\documentclass[journal,twocolumn,twoside]{IEEEtran}

\usepackage{graphics} 
\usepackage{graphicx}
\usepackage{subfigure}
\usepackage{amssymb}  
\usepackage{cite}
\usepackage{amsmath}
\usepackage{algorithm}
\usepackage{algorithmic}
\usepackage{epstopdf}
\usepackage{booktabs}
\usepackage{multirow}
\usepackage{etoolbox}
\usepackage{mathdots}
\usepackage{xcolor, float}
\usepackage{hyperref}
\makeatletter
\patchcmd{\@makecaption}
  {\scshape}
  {}
  {}
  {}
\makeatother
\usepackage{diagbox}
\usepackage{bm}
\usepackage{subfigure}
\usepackage{lipsum}
\usepackage{braket,amsfonts,amsopn}

\def\norm #1{\left\|#1\right\|}

\def\twon #1{\left\|#1\right\|_2}

\def\frobn #1{\left\|#1\right\|_{\text{F}}}

\def\abs #1{\left|#1\right|}
\def\inp #1{\left\langle#1\right\rangle}

\def\st{\text{subject to }}

\def\bC{\mathbb{C}}

\def\bR{\mathbb{R}}
\def\bE{\mathbb{E}}

\def\bS{\mathbb{S}}

\def\m #1{\boldsymbol{#1}}

\def\cD{\mathcal{D}}

\def\cL{\mathcal{L}}

\def\cO{\mathcal{O}}

\def\cT{\mathcal{T}}

\def\bee{\begin{equation}}
\def\ene{\end{equation}}

\def\beq{\begin{eqnarray}}
\def\enq{\end{eqnarray}}
\def\lentwo{\setlength\arraycolsep{2pt}}

\newtheorem{lem}{Lemma}
\newtheorem{rem}{Remark}

\newtheorem{thm}{Theorem}
\newtheorem{prop}{Proposition}

\newenvironment{proof}{{\noindent\it \quad Proof:}\;}{\hfill $\square$\par}

\def\equ #1{\begin{equation}#1\end{equation}}
\def\equa #1{\begin{eqnarray}#1\end{eqnarray}}
\def\sbra #1{\left(#1\right)}
\def\mbra #1{\left[#1\right]}
\def\lbra #1{\left\{#1\right\}}
\def\diag #1{\text{diag}#1}
\def\tr #1{\text{tr}#1}

\def\rank #1{\text{rank}#1}
\def\st {\text{ subject to }}

\DeclareMathOperator*{\argmin}{arg\,min}

\title{A Robust and Statistically Efficient Maximum-Likelihood Method for DOA Estimation Using Sparse Linear Arrays}
\author{Zai Yang, Xinyao Chen, and Xunmeng Wu
\thanks{Part of this paper was presented in the 2021 CIE IEEE International Conference \cite{yang2021maximum} and part will be presented in the 2022 IEEE International Conference on Acoustics, Speech and Signal Processing (ICASSP) \cite{chen2022localizing}.

The authors are with the School of Mathematics and Statistics, Xi'an Jiaotong University, Xi'an 710049, China (e-mail: yangzai@xjtu.edu.cn).}}

\begin{document}
\maketitle


\begin{abstract}

A recent trend of research on direction-of-arrival (DOA) estimation is to localize more uncorrelated sources than sensors by using a proper sparse linear array (SLA) and the Toeplitz covariance structure, at a cost of robustness to source correlations. In this paper, we make an attempt to achieve the two goals simultaneously by using a single algorithm. In order to statistically efficiently localize a maximal number of uncorrelated sources, we propose an effective algorithm for the stochastic maximum likelihood (SML) method based on elegant problem reformulations and the alternating direction method of multipliers (ADMM). We prove that the SML is robust to source correlations though it is derived under the assumption of uncorrelated sources. The proposed algorithm is usable for arbitrary SLAs (e.g., minimum redundancy arrays, nested arrays and coprime arrays) and is named as {\em m}aximum-likelihood {\em e}stimation via {\em s}equential {\em A}DMM (MESA). Extensive numerical results are provided that collaborate our analysis and demonstrate the statistical efficiency and robustness of MESA among state-of-the-art algorithms.

\end{abstract}



\section{Introduction}


Direction-of-arrival (DOA) estimation is a fundamental problem in statistical and array signal processing. It refers to the problem of estimating the directions of a number of sources impinging on a sensor array given a series of snapshots of the output of the sensor array \cite{stoica2005spectral}. In this paper, we consider DOA estimation for far-field narrowband sources using a uniform or sparse linear array (ULA or SLA), resulting in a DOA estimation problem equivalent to multiple-snapshot spectral analysis, a topic at the core of wireless channel estimation \cite{barbotin2012estimation}, radar signal processing\cite{li2007mimo}, structural health monitoring\cite{heylen2006modal} and fluorescence microscopy \cite{rust2006sub}. The SLA corresponds to the missing data case in the language of spectral analysis \cite{wang2006spectral} or compressive data in the language of compressed sensing \cite{candes2006robust, tang2013compressed} and brings new challenges to theoretical analysis and algorithm design.

The use of SLAs for DOA estimation dates back to \cite{moffet1968minimum} and has been extensively studied in the past two decades with an emphasis of localizing $\cO(M^2)$ sources using $M$ sensors only. The key to achieving such a goal is that under the assumption of uncorrelated sources the data covariance matrix regarding a ULA becomes Toeplitz and thus can be determined by a few of its entries. With this in mind, different array geometries for SLAs, e.g.~minimum redundancy arrays (MRAs) \cite{moffet1968minimum}, nested arrays \cite{pal2010nested,liu2016super1,shi2018generalized} and coprime arrays \cite{vaidyanathan2011sparse,qin2015generalized}, have been proposed that determine which entries (indexed by the coarray) are sampled to reconstruct the whole or a shrunk version of the Toeplitz covariance matrix. The assumption of uncorrelated sources is crucial to guarantee the Toeplitz covariance structure, however, it is not always satisfied. In fact, correlated and coherent (fully correlated) sources usually occur in practice due to multipath propagations and other effects, and dealing with them has always been a central topic in DOA estimation (see, e.g., \cite{shan1985spatial,pillai1989performance, tan1997estimating, liu2012spatial, tao2014two,qin2017doa}). Consequently, the goal of localizing more uncorrelated sources than sensors by using the Toeplitz covariance structure seemingly contradicts with the one of robust localization of highly correlated and coherent sources, making the practical use of previous methods questionable in correlated environments. In the present work, we make the first attempt to achieve the two goals simultaneously and resolve the above concern.

It is well-known that the maximum likelihood (ML) method, if solvable, provides benchmark performance for DOA estimation. Under the assumption of uncorrelated sources, the stochastic ML (SML) method can be used to localize the maximal number of sources with statistical efficiency. Its asymptotic performance, in terms of the Cram\'{e}r-Rao bound (CRB), has been well understood \cite{wang2016coarrays,liu2017cramer}. However, few algorithms have been proposed for the SML method since it resorts to a highly nonconvex optimization problem. The challenges arise due to the nonlinearity with respect to the DOAs, the nonconvex log-det term in the SML criterion function and the source number constraint, and it becomes even worse in the SLA case. In this work, in order to achieve the aforementioned two goals simultaneously, we present an effective algorithm for the SML and prove that the SML for uncorrelated sources is robust to source correlations. Our main contributions are summarized below.

\begin{enumerate}
\item We start with the specialized ULA case and formulate the SML optimization problem as a rank-constrained Toeplitz covariance estimation problem, which is further transformed as sequential rank-constrained semidefinite programs (SDPs) by applying a majorization-minimization (MM) technique \cite{sun2016majorization}. An elegant reformulation of the rank-constrained SDP is derived to fit and solved using the alternating direction method of multipliers (ADMM) encouraged by its successes in solving nonconvex problems \cite{boyd2011distributed,diamond2018general}. The resulting algorithm is named as {\em m}aximum-likelihood {\em e}stimation via {\em s}equential {\em A}DMM (MESA) (see Section \ref{sec:ULA}).
\item In the general SLA case, we repeat the above derivations and show that the SML problem can be similarly solved, extending MESA to this case (see Section \ref{sec:SLA}).
\item While the SML method concerned in the present paper is derived under the assumption of uncorrelated sources, we prove that it produces consistent estimates of the DOAs and the source powers (regardless of source correlations) as the noise vanishes, implying its robustness to correlated and coherent sources (see Section \ref{sec:robust}).
\item Numerical results are provided confirming statistical efficiency of MESA for uncorrelated source localization, in cases when the source number is less than or greater than the sensor number, and its robustness to correlated and coherent sources (see Section \ref{sec:simulation}).
\end{enumerate}

\subsection{Relations to Prior Art}
The SML method is known also as unconditional ML and has a long history of research. Its asymptotic performance in the ULA case, in terms of the CRB, is well documented in literature \cite{stoica1990performance,stoica2001stochastic}. To solve the SML optimization problem, expectation maximization (EM) and Newton-type algorithms have been proposed in earlier works \cite{feder1988parameter,starer1992newton} but their performance heavily depends on the initialization step. Instead of solving for an exact ML estimator, great efforts have been made to develop algorithms that have the same asymptotic performance as the SML. Such examples include multiple signal classification (MUSIC) \cite{schmidt1986multiple}, method of direction estimation (MODE) \cite{stoica1990maximum} and weighted subspace fitting (WSF) \cite{viberg1991detection}. Good reviews can be found in \cite{ottersten1993exact,krim1996two}. But it is worth noting that these algorithms usually consider the ULA and assume deterministic (as opposed to uncorrelated) sources and thus cannot be used to localize more sources than sensors with an SLA.

Sparse optimization and compressed sensing methods \cite{gorodnitsky1997sparse, malioutov2005sparse,candes2006stable}, which have become popular since early of this century, do not use explicitly the array geometry and fit into the SLA case. The recent atomic norm and gridless compressed sensing methods \cite{fang2014super,yang2016exact, fernandez2016super, yang2016enhancing,zhu2019multi} are remedies of earlier compressed sensing methods by working with continuous (as opposed to on-grid) DOAs and providing theoretical guarantees. These methods do not make statistical assumptions on the sources and are robust to source correlations. But correspondingly, they cannot localize more sources than sensors. Readers are referred to \cite{yang2018sparse} for a review.

To localize more uncorrelated sources than sensors, a coarray-based averaging/selection (CBA/S) step is usually adopted to explicitly use the Toeplitz covariance structure and transform the sample covariance matrix regarding the SLA as an output regarding an enlarged virtual ULA, followed by a DOA estimation method for ULAs; see such two-step estimation approaches in \cite{pal2011coprime,liu2015remarks, zhang2013sparsity,tan2014direction,shen2016underdetermined, zhou2018direction}, to name just a few. It is shown in \cite{wang2016coarrays} that CBA combined with spatial-smoothing (SS) MUSIC results in strictly non-efficient solutions. A state-of-the-art method is proposed in \cite{sedighi2019asymptotically} that uses a weighted least square (WLS) criterion for the vectorized sample covariance and is shown to yield an asymptotically efficient estimator. An iterative algorithm is also proposed to solve the resulting nonconvex optimization problem. While these methods are tailored for uncorrelated sources, it is confirmed by numerical results in this paper that they are indeed sensitive to highly correlated sources. In contrast to this, MESA achieves statistical efficiency and robustness to source correlations simultaneously.

Several algorithms have been proposed to deal with a mixture of uncorrelated and coherent sources given the source coherence structure \cite{xu2006deflation,liu2012spatial,tao2014two,qin2017doa}. Differently from these algorithms, MESA allows the sources to be correlated but noncoherent and needs only the total number of sources (as opposed to the detailed coherence structure).

The SML method in the ULA case is closely related to structured (to be specific, Toeplitz) covariance estimation (see e.g., \cite{burg1982estimation,abramovich1998positive1, li1999computationally,romero2015compressive}) because the data covariance matrix is the sum of a low-rank Toeplitz covariance and the noise covariance, where the rank is specified by the source number and the DOAs are uniquely determined by the Toeplitz covariance matrix. While the low-rank constraint is a major challenge for Toeplitz covariance estimation, it is explicitly considered in \cite{kang2015computationally, babu2016melt}. In \cite{kang2015computationally}, the Toeplitz structure is relaxed initially and then used to obtain a Toeplitz approximation of an intermediate solution, which does not result in an exact SML estimator. In \cite{babu2016melt}, the noise variance is assumed known and the Carath\'{e}odory-Fej\'{e}r theorem \cite[Theorem 11.5]{yang2018sparse} is invoked to approximate the Toeplitz covariance by a Vandermonde decomposition in which the frequency nodes of the Vandermonde matrix are restricted on a fixed grid so that the original problem is transformed as one of nonnegative sparse vector recovery. In contrast to these methods, we make no approximations or relaxations and MESA solves the exact SML. Moreover, MESA is usable in the SLA case.

Since the difficulty in solving the SML problem partly comes from the source number constraint, which is known as signal sparsity in compressed sensing, it is relaxed in sparse Bayesian learning (SBL) methods \cite{wipf2007empirical,liu2012efficient,das2017narrowband}. Similar relaxation techniques are also used in covariance fitting methods \cite{ottersten1998covariance,stoica2011spice,yang2014discretization, qiao2017gridless,wu2017toeplitz}, which are approximate versions of the SML method by using convex surrogates for its criterion function. Interestingly, it has been empirically observed in \cite{wipf2007empirical,stoica2011spice,yang2014discretization} that the resulting algorithms are robust to source correlations though they are derived by assuming uncorrelated sources. In the recent work \cite{pote2020robustness}, the case of two correlated sources is considered and it is shown that if the DOAs and the noise variance are known {\em a priori}, then the source powers can be stably estimated from the SML method. In contrast to this, our result on robustness is applicable to any source number and shows that the DOAs can be accurately estimated jointly with the source powers, at least in the high SNR regime. It also partially explains the observations in \cite{wipf2007empirical,stoica2011spice,yang2014discretization}.

\subsection{Notation}
The sets of real and complex numbers are denoted by $\bR$ and $\bC$ respectively. For vector $\m{x}$, $\diag\sbra{\m{x}}$ denotes a diagonal matrix with $\m{x}$ on the diagonal. The $ j $th entry of vector $  \boldsymbol{x} $ is $ x_j $. For matrix $ \boldsymbol{A}$,  $ \boldsymbol{A}^{T} $, $ \boldsymbol{A}^{H} $, $ | \boldsymbol{A}| $, $ \boldsymbol{A}^{-1} $, $\rank\sbra{\m{A}}$, $\tr\sbra{\m{A}}$ and $\frobn{\m{A}}$ denote the matrix transpose, conjugate transpose, determinant, inverse, rank, trace and Frobenius norm of $ \boldsymbol{A}$, respectively. The complex conjugate of scale $x$ is denoted by $\overline{x}$. The notation $\boldsymbol{A}\geq 0 $ means that $\boldsymbol{A} $ is Hermitian positive semidefinite. For index set $\Omega$ and matrix $\m{A}$, $\m{A}_{\Omega}$ represents a submatrix of $\m{A}$ obtained by keeping only the rows indexed by $\Omega$ unless otherwise stated. The inner product is represented by $\inp{\cdot,\cdot}$. For matrices $\m{Y}$ and $\m{C}\geq \m{0}$, we define
\equ{\tr\sbra{\m{Y}^H\m{C}^{-1}\m{Y}} = \min_{\m{X}} \tr\sbra{\m{X}}, \st \begin{bmatrix} \m{X} & \m{Y}^H \\ \m{Y} & \m{C} \end{bmatrix} \geq \m{0}}
whenever $\m{C}$ is positive definite or not. The expectation of a random variable is denoted by $\bE[\cdot]$.

\section{Preliminaries} \label{sec:preliminary}

\subsection{DOA Estimation Using SLAs}
An $M$-element SLA of aperture $N-1$ composes a subset of an $N$-element virtual ULA. Let the index set $\Omega\subset\lbra{1,\dots,N}$, of cardinality $M\leq N$, denote the SLA. We first consider the specialized ULA case when $\Omega=\lbra{1,\dots,N}$ and $M=N$. Assume that $K$ far-field narrowband sources impinge on the ULA in which adjacent sensors are placed by half a wavelength apart. The output of the sensor array at each snapshot composes an $N\times 1$ complex vector $\m{y}$ that can be modeled as \cite{stoica2005spectral,krim1996two}:
\begin{equation}\label{eq:model1}
\boldsymbol{y}(l)=\sum_{k=1}^{K} \boldsymbol{a}\left(f_{k}\right) x_{k}(l)+\boldsymbol{e}(l), \quad l=1,\dots,L,
\end{equation}
where $L$ is the number of snapshots, $x_k(l)$ is the $k$th (complex) source signal at the $l$th snapshot, $f_k\in[-\frac{1}{2},\frac{1}{2})$ has a one-to-one connection to the $k$th DOA $\theta_k\in [-90^\circ, 90^\circ)$ by $f_k = \frac{1}{2}\sin \theta_k$, $\m{a}\sbra{f_k}$ denotes an $N\times 1$ steering vector given by
\equ{\m{a}\sbra{f} = [1, e^{i2\pi f}, \ldots, e^{i2(N-1)\pi f}]^{T}, }
and $\boldsymbol{e}(l)$ is the vector of complex noise. It is seen that all snapshots share the same parameters $\lbra{\theta_k}$ and $\lbra{f_k}$. By stacking $\lbra{x_k(l)}, \lbra{f_k}$ into vectors $\m{x}(l), \m{f}$ and defining the steering matrix $\m{A}\sbra{\m{f}} = \mbra{\m{a}\sbra{f_1}, \ldots, \m{a}\sbra{f_K}}$ that is $N\times K$ Vandermonde, the data model in \eqref{eq:model1} is written compactly as:
\begin{equation}\label{eq:model}
\boldsymbol{y}(l)=\boldsymbol{A}(\boldsymbol{f}) \boldsymbol{x}(l)+\boldsymbol{e}(l), \quad l=1,\dots,L.
\end{equation}

In the general SLA case, the array output at one snapshot is a subvector of $\boldsymbol{y}(l)$, denoted by $\boldsymbol{y}_{\Omega}(l)$. The data model in \eqref{eq:model} thus becomes
\begin{equation}\label{eq:model_sla}
\boldsymbol{y}_{\Omega}(l)=\boldsymbol{A}_{\Omega}(\boldsymbol{f}) \boldsymbol{x}(l)+\boldsymbol{e}_{\Omega}(l), \quad l=1,\dots,L,
\end{equation}
which encompasses \eqref{eq:model} as a special case.

Our objective is to estimate the DOAs $\lbra{\theta_k}_{k=1}^K$, or equivalently $\lbra{f_k}_{k=1}^K$, given the multiple-snapshot data $\lbra{\m{y}_{\Omega}(l)}_{l=1}^K$ under certain statistical assumptions on the source signals $\lbra{\m{x}(l)}$ and noise $\lbra{\m{e}(l)}$. Since each $f_k$ is the frequency of a sinusoid, the DOA estimation problem that we concern is equivalent to multiple-snapshot spectral estimation with missing data. We focus on the estimation of $\lbra{f_k}_{k=1}^K$ throughout this paper.

\subsection{The SML Method for DOA Estimation} \label{sec:sml}
We make the following assumptions to derive the SML method for DOA estimation.
\begin{itemize}
\item[A1:] The sources $\lbra{\m{x}(l)}$ are spatially and temporally independent and follow a complex Gaussian distribution with zero mean and covariance $\m{P}= \diag\sbra{p_1,\dots,p_K}$, where $p_k>0$ denotes the $k$th sources power;
\item[A2:] The noises $\lbra{\m{e}(l)}$ are spatially and temporally independent and each entry follows a complex Gaussian distribution with zero mean and variance $\sigma>0$;
\item[A3:] The sources and noises are independent.
\end{itemize}
It follows immediately that $\lbra{\m{y}_{\Omega}(l)}$ are i.i.d.~Gaussian with zero mean and covariance
\equ{\m{R}_{\Omega} = \m{A}_{\Omega}(\m{f})\m{P}\m{A}_{\Omega}^H(\m{f}) + \sigma\m{I}. \label{eq:Rinfp}}
By maximizing the likelihood criterion, or equivalently minimizing the negative log-likelihood function, we obtain the SML optimization problem as:
\equ{\min_{\m{f}, \m{p},\sigma} \ln\abs{\m{R}_{\Omega}} + \tr\sbra{\m{R}_{\Omega}^{-1}\widehat{\m{R}}_{\Omega}}, \label{eq:sml_sla}}
where
\equ{\widehat{\m{R}}_{\Omega} = \frac{1}{L}\sum_{l=1}^L\m{y}_{\Omega}(l)\m{y}_{\Omega}^H(l)}
is the sample covariance matrix.

The SML method has good statistical properties. But the SML problem in \eqref{eq:sml_sla} is nonconvex and complicated to solve due to the log-det term and the nonlinearity of $\m{R}_{\Omega}$ with respect to $\lbra{f_k}_{k=1}^K$. Moreover, the SML is derived under the assumption of uncorrelated sources and its performance is unclear in presence of source correlations.

%

\subsection{The ADMM Algorithm}
The ADMM algorithm solves the following optimization problem:
\equ{\min_{\m{x}\in\cD_{1},\m{q}\in\cD_{2}} g(\m{x}) + h(\m{q}), \st \m{A}\m{x} + \m{B}\m{q} = \m{c}, \label{eq:admm_prb}}
where $\cD_{1}, \cD_{2}$ defines the feasible domain of $\m{x},\m{q}$ respectively. Write the augmented Lagrangian function as:
\equ{\begin{split}
&\cL_{\mu}\sbra{\m{x},\m{q}, \m{\lambda}} \\
&= g(\m{x}) + h(\m{q}) + \inp{\m{A}\m{x} + \m{B}\m{q} - \m{c}, \m{\lambda}} + \frac{\mu}{2}\twon{\m{A}\m{x} + \m{B}\m{q} - \m{c}}^2 \\
&= g(\m{x}) + h(\m{q}) + \frac{\mu}{2}\twon{\m{A}\m{x} + \m{B}\m{q} - \m{c} + \mu^{-1}\m{\lambda}}^2 + C, \end{split}}
where $\m{\lambda}$ is a Lagrangian multiplier, $\mu>0$ is a penalty coefficient and $C$ is a constant independent of $\m{x},\m{q}$. ADMM consists of the iterations:
{\lentwo\equa{\m{x}
&\leftarrow& \argmin_{\m{x}\in\cD_1} \cL_{\mu}\sbra{\m{x},\m{q}, \m{\lambda}}, \label{eq:subpx}\\ \m{q}
&\leftarrow& \argmin_{\m{q}\in\cD_2} \cL_{\mu}\sbra{\m{x},\m{q}, \m{\lambda}}, \label{eq:subpq} \\ \m{\lambda}
&\leftarrow& \m{\lambda} + \mu\sbra{\m{A}\m{x} + \m{B}\m{q} - \m{c}}, \label{eq:updatelamda}
}}where the latest values of the other variables are always used. The ADMM algorithm has been extensively studied and practically used due to its global optimality in solving convex problems, simplicity in dealing with nonsmooth functions, and good scalability for solving high-dimensional problems \cite{boyd2011distributed}. Good performance has also been frequently achieved for nonconvex problems; see \cite{diamond2018general} and references therein. See also \cite{hong2016convergence,wang2018convergence,wang2019global} for theoretical progresses on this topic. It is worth noting that the key to using ADMM to solve a specific problem is to provide an elegant problem formulation within the ADMM framework so that the two subproblems in \eqref{eq:subpx} and \eqref{eq:subpq} can be simply and efficiently solved.

\section{MESA in the ULA Case}\label{sec:ULA}
In this section, we derive the MESA algorithm for the SML optimization problem in \eqref{eq:sml_ula} in the specialized ULA case. In this case, we write the data and sample covariance matrices $\m{R}_{\Omega}, \widehat{\m{R}}_{\Omega}$ into $\m{R}, \widehat{\m{R}}$ for simplicity and the problem to solve becomes:
\equ{\min_{\m{f}, \m{p},\sigma} \ln\abs{\m{R}} + \tr\sbra{\m{R}^{-1}\widehat{\m{R}}}, \label{eq:sml_ula}}
where
\equ{\m{R} = \m{A}(\m{f})\m{P}\m{A}^H(\m{f}) + \sigma\m{I}. \label{eq:Rinfp}}
The MESA algorithm consists of re-parameterization, majorization-minimization, problem reformulation and ADMM steps which are detailed below.

\subsection{Re-parameterization}
The data covariance matrix $\m{R}$ is a highly nonlinear function of $\lbra{f_k}$. To overcome such nonlinearity, a common scheme is to utilize the fact that the first term $\m{A}(\m{f})\m{P}\m{A}^H(\m{f})$ in \eqref{eq:Rinfp} is rank-$K$ positive-semidefinite Hermitian Toeplitz and do the re-parameterization:
	\equ{\m{R} = \cT\m{t} + \sigma\m{I},\quad \cT\m{t}\geq \m{0},\quad \rank\sbra{\cT\m{t}} = K, \label{eq:R}}
	where $ \mathcal{T} \boldsymbol{t}=\left(t_{i-j}\right)_{N \times N} $ with $ \boldsymbol{t}=\left[t_{1-N}, \ldots, t_{N-1}\right]^{T} $ and $ t_{-j}=\bar{t}_{j}, j=0, \ldots, N-1 $. It follows from the Carath\'{e}odory-Fej\'{e}r theorem \cite[Theorem 11.5]{yang2018sparse} that the $\cT\m{t}$ above has a one-to-one connection to $\lbra{\m{f}, \m{p}}$ given $K<N$. Consequently, the original SML problem \eqref{eq:Rinfp} is transformed into a rank-constrained Toeplitz covariance estimation problem in which $\m{R}$ is a linear function of the variables $\lbra{\m{t},\sigma}$.
Once $\m{t}$ is solved for, the variables $\lbra{\m{f}, \m{p}}$ can be computed from $\cT\m{t}$ by a subspace method such as root-MUSIC \cite{barabell1983improving}.
	
\subsection{Majorization Minimization}
The objective function in \eqref{eq:sml_ula} is nonconvex with respect to $\m{R}$ since the log-det function $ \ln |\boldsymbol{R}|$ is concave
on the positive semidefinite cone. A commonly used locally convergent method is the majorization-minimization (MM) algorithm (see, e.g., \cite{fazel2003log}) that drives the objective function downhill by minimizing a simple surrogate function. At the $j$th iteration of MM, the SML objective function is linearized (and thus majorized) at the previous iterate $\m{R}_{j-1}=\cT\m{t}_{j-1} + \sigma_{j-1}\m{I}$, yielding the problem (by omitting constant terms):
	\begin{equation}\label{surrogate function}
		\min \operatorname{tr}\left(\boldsymbol{R}_{j-1}^{-1} \boldsymbol{R}\right)+\operatorname{tr}\left(\boldsymbol{R}^{-1} \widehat{\boldsymbol{R}}\right).
	\end{equation}
Substituting \eqref{eq:R} into \eqref{surrogate function}, we obtain the problem to solve at the $j$th iteration as:
\begin{equation}
		\begin{array}{l}
			\min\limits _{\m{t}, \sigma\geq0} \operatorname{tr}(\boldsymbol{R}_{j-1}^{-1}(\cT\m{t}+\sigma \boldsymbol{I}))+\operatorname{tr}\left((\cT\m{t}+\sigma \boldsymbol{I})^{-1} \widehat{\m{R}}\right), \\
			\text { subject to } \cT\m{t}\in \mathbb{S}_{+}^{K},
		\end{array} \label{eq:sml_j}
\end{equation}
where $\mathbb{S}_{+}^{K}$ is the set of positive semidefinite matrices of rank no greater than $ K $.


\subsection{Problem Reformulation}
Let $ \boldsymbol{W}=\boldsymbol{R}_{j-1}^{-1} $ and $ \widehat{\boldsymbol{Y}} $ be any matrix satisfying that
\bee\label{Yhat}
	\widehat{\boldsymbol{R}}= \widehat{\boldsymbol{Y}}\widehat{\boldsymbol{Y}}^{H},
	\ene
where $\widehat{\boldsymbol{Y}}$ has at most $\min(L,N)$ columns.
	The objective function in \eqref{eq:sml_j} then becomes:
	\begin{equation}\label{min C sigma}
		\operatorname{tr}(\boldsymbol{W}(\cT\m{t}+\sigma \boldsymbol{I}))+\operatorname{tr}\left(\widehat{\boldsymbol{Y}}^{H}(\cT\m{t}+\sigma \boldsymbol{I})^{-1} \widehat{\boldsymbol{Y}}\right).
	\end{equation}
Making use of the following identity  \cite[Lemma 5]{yang2015gridless}:
	\begin{equation}\label{PA 3}
		\begin{array}{l}
			\operatorname{tr}\left(\boldsymbol{X}^{H}\left(\boldsymbol{R}_{1}+\boldsymbol{R}_{2}\right)^{-1} \boldsymbol{X}\right) \\
			=\min \limits_{\boldsymbol{Z}} \operatorname{tr}\left(\boldsymbol{Z}^{H} \boldsymbol{R}_{1}^{-1} \boldsymbol{Z}\right)+\operatorname{tr}\left((\boldsymbol{X}-\boldsymbol{Z})^{H} \boldsymbol{R}_{2}^{-1}(\boldsymbol{X}-\boldsymbol{Z})\right),
		\end{array}
	\end{equation}
where $ \boldsymbol{R}_{1}, \boldsymbol{R}_{2} \geq 0 $, the function in \eqref{min C sigma} becomes a function of $ \m{t}, \sigma, \boldsymbol{Z} $:
	\begin{equation}\label{PA 4}
		\operatorname{tr}(\boldsymbol{W}(\cT\m{t}+\sigma \boldsymbol{I}))+\operatorname{tr}\left(\boldsymbol{Z}^{H} \mbra{\cT\m{t}}^{-1} \boldsymbol{Z}\right)+\sigma^{-1}\|\widehat{\boldsymbol{Y}}-\boldsymbol{Z}\|_{\mathrm{F}}^{2}.
	\end{equation}
Since in \eqref{PA 4} the optimizer to $ \sigma $ is given in close-form by:
	\begin{equation}\label{PA 6}		\sigma^{*}=\frac{1}{\sqrt{\operatorname{tr}(\boldsymbol{W})}}\| \widehat{\boldsymbol{Y}}-\boldsymbol{Z}\|_{\mathrm{F}},
\end{equation}
the objective in \eqref{PA 4} can be concentrated with respect to $\m{t}, \boldsymbol{Z} $, yielding the following problem to solve:
\begin{equation}\label{min C Z}
		\begin{array}{l}
			\min \limits_{\m{t}, \boldsymbol{Z}} \operatorname{tr}(\boldsymbol{W} \cT\m{t})+\operatorname{tr}\left(\boldsymbol{Z}^{H} \mbra{\cT\m{t}}^{-1} \boldsymbol{Z}\right)+2 \sqrt{\operatorname{tr}(\boldsymbol{W})}\|\widehat{\boldsymbol{Y}}-\boldsymbol{Z}\|_{\mathrm{F}}, \\
           \text { subject to } \cT\m{t} \in \bS_+^K,
		\end{array}
	\end{equation}
	or equivalently,
	\begin{equation}\label{min t x z 1}
		\begin{array}{l}
			\min \limits_{\boldsymbol{t}, \boldsymbol{X}, \boldsymbol{Z}} \operatorname{tr}(\boldsymbol{W} \mathcal{T} \boldsymbol{t})+\operatorname{tr}(\boldsymbol{X})+2 \sqrt{\operatorname{tr}(\boldsymbol{W})}\|\widehat{\boldsymbol{Y}}-\boldsymbol{Z}\|_{\mathrm{F}}, \\
			\text { subject to }\begin{bmatrix}
				\boldsymbol{X} & \boldsymbol{Z}^{H} \\
				\boldsymbol{Z} & \mathcal{T} \boldsymbol{t}
			\end{bmatrix} \geq \mathbf{0}, \rank\sbra{\mathcal{T} \boldsymbol{t}} \leq K.
		\end{array}
	\end{equation}

We next show that the problem in \eqref{min t x z 1} is equivalent to the following:
	\begin{equation}\label{min t x z 2}
		\begin{array}{l}
			\min\limits _{\boldsymbol{t}, \boldsymbol{X}, \boldsymbol{Z}} \operatorname{tr}(\boldsymbol{W} \mathcal{T} \boldsymbol{t})+\operatorname{tr}(\boldsymbol{X})+2 \sqrt{\operatorname{tr}(\boldsymbol{W})}\|\widehat{\boldsymbol{Y}}-\boldsymbol{Z}\|_{\mathrm{F}}, \\
			\text { subject to }\begin{bmatrix}
				\boldsymbol{X} & \boldsymbol{Z}^{H} \\
				\boldsymbol{Z} & \mathcal{T} \boldsymbol{t}
			\end{bmatrix}\geq \mathbf{0}, \rank\begin{bmatrix}
				\boldsymbol{X} & \boldsymbol{Z}^{H} \\
				\boldsymbol{Z} & \mathcal{T} \boldsymbol{t}
			\end{bmatrix} \leq K.
		\end{array}
	\end{equation}
	In particular, it is easy to see that the constraint $  \rank(\mathcal{T} \boldsymbol{t}) \leq K $ in \eqref{min t x z 1} is
	implied by $  \rank\begin{bmatrix}
		\boldsymbol{X} & \boldsymbol{Z}^{H} \\
		\boldsymbol{Z} & \mathcal{T} \boldsymbol{t}
	\end{bmatrix} \leq K $ in \eqref{min t x z 2}. To show the
	equivalence between  \eqref{min t x z 1} and \eqref{min t x z 2}, it suffices to show that the latter constraint is feasible for any optimizer to \eqref{min t x z 1}. Denote by
	$ (\boldsymbol{t}^{\ast}, \boldsymbol{X}^{\ast}, \boldsymbol{Z}^{\ast}) $ an optimizer to \eqref{min t x z 1} and suppose $  \mathcal{T} \boldsymbol{t}^{\ast}= \boldsymbol{T} \boldsymbol{T}^{H} $ where $ \boldsymbol{T} $ is an $ L\times \bar{K} $ matrix with full column rank $ \bar{K}\leq K $.
	Since $ \begin{bmatrix}
		\boldsymbol{X}^{*} & \boldsymbol{Z}^{* H} \\
		\boldsymbol{Z}^{*} & \mathcal{T} \boldsymbol{t}^{*}
	\end{bmatrix} \geq 0 $, we have $ \boldsymbol{Z}^{\ast}=\boldsymbol{T}\boldsymbol{V} $ for some $ \boldsymbol{V} $ due to the column/row inclusion property
	and $ \boldsymbol{X}^{\ast}= \boldsymbol{Z}^{\ast H}( \mathcal{T} \boldsymbol{t}^{*})^{\dagger} \boldsymbol{Z}^{\ast}= \boldsymbol{V}^{H} \boldsymbol{V} $ . Therefore,
	\begin{equation}\label{PA 18}
		\begin{bmatrix}
			\boldsymbol{X}^{*} & \boldsymbol{Z}^{* H} \\
			\boldsymbol{Z}^{*} & \mathcal{T} \boldsymbol{t}^{*}
		\end{bmatrix}=\begin{bmatrix}
			\boldsymbol{V}^{H} \boldsymbol{V} & \boldsymbol{V}^{H} \boldsymbol{T}^{H} \\
			\boldsymbol{T} \boldsymbol{V} & \boldsymbol{T} \boldsymbol{T}^{H}
		\end{bmatrix}=\begin{bmatrix}
			\boldsymbol{V}^{H} \\
			\boldsymbol{T}
		\end{bmatrix}\begin{bmatrix}
			\boldsymbol{V}^{H} \\
			\boldsymbol{T}
		\end{bmatrix}^{H}
	\end{equation}
	whose rank is $ \bar{K}\leq K $, completing the proof.

\subsection{Using ADMM}
We introduce an auxiliary matrix variable $ \boldsymbol{Q} $ and rewrite \eqref{min t x z 2} as
\begin{equation}\label{min t x z Q}
\begin{split}
&\min_{\m{t},\m{X},\m{Z},\m{Q}\in\bS_+^K} \operatorname{tr}(\boldsymbol{W} \mathcal{T} \boldsymbol{t})+\operatorname{tr}(\boldsymbol{X})+2 \sqrt{\operatorname{tr}(\boldsymbol{W})}\|\widehat{\boldsymbol{Y}}-\boldsymbol{Z}\|_{\mathrm{F}},\\ &\st \m{Q} = \begin{bmatrix} \m{X} & \m{Z}^H \\ \m{Z} & \cT\m{t} \end{bmatrix}, \end{split}
\end{equation}
which is exactly in the form of \eqref{eq:admm_prb} by identifying that $\m{x}=\lbra{\m{t},\m{X},\m{Z}}$, $\m{q}=\m{Q}$, $g(\m{x}) = \operatorname{tr}(\boldsymbol{W} \mathcal{T} \boldsymbol{t})+\operatorname{tr}(\boldsymbol{X})+2 \sqrt{\operatorname{tr}(\boldsymbol{W})}\|\widehat{\boldsymbol{Y}}-\boldsymbol{Z}\|_{\mathrm{F}}$, $h(\m{q})=0$ and $\cD_2 = \bS_+^K$. Following the procedures of ADMM, we introduce the Hermitian Lagrangian multiplier $\boldsymbol{\Lambda}$ and write the augmented Lagrangian function as:
\begin{equation}\label{Lagrangian function}
\begin{split}
&\mathcal{L}_{\mu}(\m{t},\m{X},\m{Z},\m{Q}) \\
&=\operatorname{tr}(\boldsymbol{W} \mathcal{T} \boldsymbol{t})+\operatorname{tr}(\boldsymbol{X})+2 \sqrt{\operatorname{tr}(\boldsymbol{W})}\|\widehat{\boldsymbol{Y}}-\boldsymbol{Z}\|_{\mathrm{F}}\\
&\quad +\operatorname{tr}((\boldsymbol{Q}-\mathcal{A} \boldsymbol{x}) \boldsymbol{\Lambda})+\frac{\mu}{2}\|\boldsymbol{Q}-\mathcal{A} \boldsymbol{x}\|_{\mathrm{F}}^{2} \\
&=\operatorname{tr}(\boldsymbol{W} \mathcal{T} \boldsymbol{t})+\operatorname{tr}(\boldsymbol{X})+2 \sqrt{\operatorname{tr}(\boldsymbol{W})}\|\widehat{\boldsymbol{Y}}-\boldsymbol{Z}\|_{\mathrm{F}}\\
&\quad +\frac{\mu}{2}\left\|\boldsymbol{Q}-\mathcal{A} \boldsymbol{x}+\mu^{-1} \boldsymbol{\Lambda}\right\|_{\mathrm{F}}^{2}-\frac{1}{2 \mu}\|\boldsymbol{\Lambda}\|_{\mathrm{F}}^{2}.
\end{split}
\end{equation}
The remaining task is to solve the two subproblems in \eqref{eq:subpx} and \eqref{eq:subpq}.

To solve \eqref{eq:subpx}, by partitioning $ \boldsymbol{Q}=\begin{bmatrix}
		\boldsymbol{Q}_{1} & \boldsymbol{Q}_{2}^{H} \\
		\boldsymbol{Q}_{2} & \boldsymbol{Q}_{3}
	\end{bmatrix}$ and $ 	\boldsymbol{\Lambda}=\begin{bmatrix}
		\boldsymbol{\Lambda}_{1} & \boldsymbol{\Lambda}_{2}^{H} \\
		\boldsymbol{\Lambda}_{2} & \boldsymbol{\Lambda}_{3}
	\end{bmatrix} $
as
	$ \begin{bmatrix}
		\boldsymbol{X} & \boldsymbol{Z}^{H} \\
		\boldsymbol{Z} & \mathcal{T} \boldsymbol{t}
	\end{bmatrix} $, we note that the objective function $\mathcal{L}_{\mu}$ is separable in $\lbra{\m{t},\m{X},\m{Z}}$ and thus they can be solved for separately. To solve for $\m{t}$, we equate the derivative of $\mathcal{L}_{\mu}$ with respect to $ \boldsymbol{t}$ to zero and obtain the update:
\begin{equation}\label{t update}
		\boldsymbol{t}  \leftarrow \left(\mathcal{T}^{*} \mathcal{T}\right)^{-1} \mathcal{T}^{*}\left(\boldsymbol{Q}_{3}+\mu^{-1} \boldsymbol{\Lambda}_{3}-\mu^{-1} \boldsymbol{W}\right),
	\end{equation}
where $\cT^*$ is the Hermitian adjoint of $\cT$.
Similarly, we have that
\equ{\label{x update}
		\boldsymbol{X} \leftarrow \boldsymbol{Q}_{1}+\mu^{-1} \boldsymbol{\Lambda}_{1}-\mu^{-1} \boldsymbol{I}.}
For $\boldsymbol{Z}$, the optimization problem to solve is given by:
	\begin{equation}\label{min z}
		\min _{\boldsymbol{Z}} \sqrt{\operatorname{tr}(\boldsymbol{W})}\|\widehat{\boldsymbol{Y}}-\boldsymbol{Z}\|_{\mathrm{F}}+\frac{\mu}{2}\left\|\boldsymbol{Q}_{2}+\mu^{-1} \boldsymbol{\Lambda}_{2}-\boldsymbol{Z}\right\|_{\mathrm{F}}^{2},
	\end{equation}
yielding the update:
	\begin{equation}\label{Z update}
		\boldsymbol{Z} \leftarrow \widehat{\boldsymbol{Y}}-\left(1-\frac{\sqrt{\operatorname{tr}(\boldsymbol{W})}}{\mu\left\| \boldsymbol{L}\right\|_{\mathrm{F}}}\right)_{+} \boldsymbol{L},
	\end{equation}
	where $ (x)_{+} \triangleq \max (x, 0) $ and $ \boldsymbol{L} \triangleq \widehat{\boldsymbol{Y}}-\boldsymbol{Q}_{2}-\mu^{-1} \boldsymbol{\Lambda}_{2} $. The detailed derivations of \eqref{Z update} are deferred to Appendix \ref{app:updateZ}.
	
Solving \eqref{eq:subpq} results in the update:
	\begin{equation}\label{Q update}
		\boldsymbol{Q} \leftarrow \mathcal{P}_{\mathbb{S}_{+}^{K}}\left(\begin{bmatrix}
			\boldsymbol{X} & \boldsymbol{Z}^{H} \\
			\boldsymbol{Z} & \mathcal{T} \boldsymbol{t}
		\end{bmatrix}-\mu^{-1} \boldsymbol{\Lambda}\right),
	\end{equation}
where $\mathcal{P}_{\mathbb{S}_{+}^{K}}$ denotes the orthogonal projection onto $\mathbb{S}_{+}^{K}$ that can be computed by the truncated eigen-decomposition by keeping only the largest $K$ positive eigenvalues and associated eigenvectors.

Finally, $\mathbf{\Lambda}$ is updated according to \eqref{eq:updatelamda} as:
	\begin{equation}\label{Lambda update}
		\mathbf{\Lambda} \leftarrow \mathbf{\Lambda}+\mu\left(\boldsymbol{Q}-\begin{bmatrix}
			\boldsymbol{X} & \boldsymbol{Z}^{H} \\
			\boldsymbol{Z} & \mathcal{T} \boldsymbol{t}
		\end{bmatrix}\right).
	\end{equation}
The ADMM algorithm runs \eqref{t update}, \eqref{x update}, \eqref{Z update},  \eqref{Q update} and  \eqref{Lambda update} iteratively.
	
The overall MESA algorithm consists of the outer MM loop and the inner ADMM loop. Its computations are dominated by the truncated eigen-decomposition and the matrix inverse to compute $\m{W}$. Consequently, MESA has a computational complexity of $\cO\sbra{N^3}$ per inner iteration that is affordable in DOA estimation where the array aperture $N$ is usually small.

	
	%
	%

\section{MESA in the SLA Case}\label{sec:SLA}
In this section, we consider the general SLA case and derive the MESA algorithm by repeating the same steps as in the previous section.

\subsection{Re-parameterization}
%

For a general SLA designated by $\Omega$, let $\m{\Gamma}\in\lbra{0,1}^{M\times N}$ be the row-selection matrix satisfying that
\equ{\m{y}_{\Omega} = \m{\Gamma}\m{y}}
for any $N\times 1$ vector $\m{y}$ and its subvector $\m{y}_{\Omega}$.
Then, we have
\equ{\m{R}_{\Omega}
= \bE \m{y}_{\Omega}\m{y}_{\Omega}^H = \m{\Gamma}\m{R}\m{\Gamma}^T, \label{eq:ROmega}}
which is an $M\times M$ principal submatrix of $\m{R}$, defined in \eqref{eq:Rinfp}, and is re-parameterized as a linear function of $\sbra{\m{t},\sigma}$ following from \eqref{eq:R}.


\subsection{Majorization Minimization}
In this case, the objective function at the $j$th iteration becomes:
	\begin{equation}\label{surrogate function_Omega}
		\begin{split}
&\operatorname{tr}\left(\boldsymbol{R}_{\Omega,j-1}^{-1} \boldsymbol{R}_{\Omega}\right)+\operatorname{tr}\left(\boldsymbol{R}_{\Omega}^{-1} \widehat{\boldsymbol{R}}_{\Omega}\right)\\
&= \operatorname{tr}\left(\m{\Gamma}^T\boldsymbol{R}_{\Omega,j-1}^{-1} \m{\Gamma} \m{R}\right)+\operatorname{tr}\left(\boldsymbol{R}_{\Omega}^{-1} \widehat{\m{Y}}_{\Omega}\widehat{\m{Y}}_{\Omega}^H\right)\\
&= \operatorname{tr}\left(\boldsymbol{W} \m{R}\right)+\operatorname{tr}\left(\widehat{\m{Y}}_{\Omega}^H\boldsymbol{R}_{\Omega}^{-1} \widehat{\m{Y}}_{\Omega}\right),
	\end{split}\end{equation}
where $\boldsymbol{R}_{\Omega,j-1}$ denotes the $(j-1)$st iterate of $\boldsymbol{R}_{\Omega}$, $\boldsymbol{W}=\m{\Gamma}^T\boldsymbol{R}_{\Omega,j-1}^{-1}\m{\Gamma}$, and $\widehat{\m{Y}}_{\Omega}$ is any matrix satisfying that $\widehat{\m{R}}_{\Omega} = \widehat{\m{Y}}_{\Omega}\widehat{\m{Y}}_{\Omega}^H$ and has at most $\min(L,M)$ columns.

\subsection{Problem Reformulation}
Making use of \cite[Lemma 6]{yang2015gridless}, we obtain
\equ{\operatorname{tr}\left(\widehat{\m{Y}}_{\Omega}^H\boldsymbol{R}_{\Omega}^{-1} \widehat{\m{Y}}_{\Omega}\right) = \min_{\widehat{\m{Y}}_{\overline{\Omega}}} \operatorname{tr}\left(\widehat{\m{Y}}^H\boldsymbol{R}^{-1} \widehat{\m{Y}}\right), \label{eq:lemma6}}
where the set $\overline{\Omega}$ denotes the complement of $\Omega$. By substituting \eqref{eq:lemma6} and \eqref{eq:R} into \eqref{surrogate function_Omega}, the objective function becomes
	\begin{equation}\label{min C sigma_Omega}
		\operatorname{tr}(\boldsymbol{W}(\cT\m{t}+\sigma \boldsymbol{I}))+\operatorname{tr}\left(\widehat{\boldsymbol{Y}}^{H}(\cT\m{t}+\sigma \boldsymbol{I})^{-1} \widehat{\boldsymbol{Y}}\right)
	\end{equation}
with respect to $\sbra{\m{t},\sigma,\widehat{\m{Y}}_{\overline{\Omega}}}$, which is exactly in the form of \eqref{min C sigma}. Consequently, the same derivations as in the ULA case can be applied, yielding the following optimization problem:
	\begin{equation}\label{min t x z 1_Omega}
		\begin{array}{l}
			\min\limits _{\boldsymbol{t}, \boldsymbol{X},\widehat{\m{Y}}_{\overline{\Omega}}, \boldsymbol{Z}} \operatorname{tr}(\boldsymbol{W} \mathcal{T} \boldsymbol{t})+\operatorname{tr}(\boldsymbol{X})+2 \sqrt{\operatorname{tr}(\boldsymbol{W})}\|\widehat{\boldsymbol{Y}}-\boldsymbol{Z}\|_{\mathrm{F}}, \\
			\text { subject to }\begin{bmatrix}
				\boldsymbol{X} & \boldsymbol{Z}^{H} \\
				\boldsymbol{Z} & \mathcal{T} \boldsymbol{t}
			\end{bmatrix} \in \bS_+^K,
		\end{array}
	\end{equation}
or equivalently,
	\begin{equation}\label{min t x z 2_Omega}
		\begin{array}{l}
			\min\limits _{\boldsymbol{t}, \boldsymbol{X},\boldsymbol{Z}} \operatorname{tr}(\boldsymbol{W} \mathcal{T} \boldsymbol{t})+\operatorname{tr}(\boldsymbol{X})+2 \sqrt{\operatorname{tr}(\boldsymbol{W})}\| \widehat{\boldsymbol{Y}}_{\Omega}-\boldsymbol{Z}_{\Omega}\|_{\mathrm{F}}, \\
			\text { subject to }\begin{bmatrix}
				\boldsymbol{X} & \boldsymbol{Z}^{H} \\
				\boldsymbol{Z} & \mathcal{T} \boldsymbol{t}
			\end{bmatrix} \in \bS_+^K
		\end{array}
	\end{equation}
by noting that the solution to $\widehat{\m{Y}}_{\overline{\Omega}}$ is exactly $\m{Z}_{\overline{\Omega}}$. In this process, as in \eqref{PA 6}, we have
\begin{equation}\label{PA 7}		\sigma^{*}=\frac{1}{\sqrt{\operatorname{tr}(\boldsymbol{W})}}\| \widehat{\boldsymbol{Y}}_{\Omega}-\boldsymbol{Z}_{\Omega}\|_{\mathrm{F}}.
\end{equation}

\subsection{Using ADMM}
The only difference between \eqref{min t x z 2_Omega} and \eqref{min t x z 2} is the inclusion of the index set $\Omega$ in the term $\| \widehat{\boldsymbol{Y}}-\boldsymbol{Z}\|_{\mathrm{F}}$. Consequently, the only difference in the ADMM algorithm occurs in the update of $\m{Z}$. In particular, the objective function to minimize regarding $\m{Z}$ changes from that in \eqref{min z} to the following:
\begin{equation}\label{min zsla}
\begin{split}
&\sqrt{\operatorname{tr}(\boldsymbol{W})}\|\widehat{\boldsymbol{Y}}_{\Omega}-\boldsymbol{Z}_{\Omega} \|_{\mathrm{F}}+\frac{\mu}{2}\left\|\boldsymbol{Q}_{2}+\mu^{-1} \boldsymbol{\Lambda}_{2}-\boldsymbol{Z}\right\|_{\mathrm{F}}^{2} \\
&= \sqrt{\operatorname{tr}(\boldsymbol{W})}\| \widehat{\boldsymbol{Y}}_{\Omega}-\boldsymbol{Z}_{\Omega} \|_{\mathrm{F}}+ \frac{\mu}{2}\left\|\mbra{\boldsymbol{Q}_{2}+\mu^{-1} \boldsymbol{\Lambda}_{2}}_{\Omega}-\boldsymbol{Z}_{\Omega}\right\|_{\mathrm{F}}^{2} \\
&\quad+ \frac{\mu}{2}\left\|\mbra{\boldsymbol{Q}_{2}+\mu^{-1} \boldsymbol{\Lambda}_{2}}_{\overline{\Omega}}-\boldsymbol{Z}_{\overline{\Omega}} \right\|_{\mathrm{F}}^{2}. \end{split}
\end{equation}
Therefore, it follows from \eqref{Z update} that
\lentwo{\equa{\label{Z1 updatesla}
\boldsymbol{Z}_{\Omega}
&\leftarrow& \widehat{\boldsymbol{Y}}_{\Omega}-\left(1-\frac{\sqrt{\operatorname{tr}(\boldsymbol{W})}}{\mu\left\| \boldsymbol{L}_{\Omega}\right\|_{\mathrm{F}}}\right)_{+} \boldsymbol{L}_{\Omega},\\ \label{Z2 updatesla} \boldsymbol{Z}_{\overline{\Omega}}
&\leftarrow& \mbra{\boldsymbol{Q}_{2}+\mu^{-1} \boldsymbol{\Lambda}_{2}}_{\overline{\Omega}},
}}where $\m{L}$ is as defined below \eqref{Z update}.

The ADMM algorithm in this case runs \eqref{t update}, \eqref{x update}, \eqref{Z1 updatesla}, \eqref{Z2 updatesla}, \eqref{Q update} and \eqref{Lambda update} iteratively. Again, the overall MESA algorithm consists of the outer MM loop and the inner ADMM loop and it is illustrated in Algorithm \ref{alg:MESA}. It has a computational complexity of $\cO(N^3)$ per inner iteration and degenerates into MESA in the previous section in the specialized ULA case.


\begin{algorithm}
	\caption{Maximum-likelihood estimation via sequential ADMM (MESA)}
	\label{MESA}
	\begin{algorithmic}[1]
		\REQUIRE SLA $\Omega$, sample covariance $\widehat{\m{R}}_{\Omega}$, source number $K$.
		\ENSURE Estimates of frequencies $\m f$, source powers $\m p$ and noise power $\sigma$.
		\STATE Calculate $\widehat{\m Y}_{\Omega}=\widehat{\m{R}}_{\Omega}^{\frac{1}{2}}$;
        \STATE Initialize $\m R_{\Omega}, \m Q, \m\Lambda, \m t, \m X, \m Z$ and calculate $\m W= \m{\Gamma}^T\boldsymbol{R}_{\Omega}^{-1} \m{\Gamma}$;
		\WHILE{not converged}
		\WHILE{not converged}
		\STATE Conduct the updates in \eqref{t update}, \eqref{x update}, \eqref{Z1 updatesla}, \eqref{Z2 updatesla}, \eqref{Q update} and \eqref{Lambda update} one after one;
		\ENDWHILE
		\STATE Conduct the update for $\sigma$ in \eqref{PA 7};
        \STATE Update $\m{R}$ in \eqref{eq:R} and $\m W= \m{\Gamma}^T\boldsymbol{R}_{\Omega}^{-1} \m{\Gamma}$;
		\ENDWHILE
		\STATE Calculate the estimates of $\m f$ and $\m p$ by computing the decomposition $\cT\m{t}=\m{A}\sbra{\m{f}}\m{P}\m{A}^H\sbra{\m{f}}$ using root-MUSIC.
	\end{algorithmic} \label{alg:MESA}
\end{algorithm}

\section{Robustness to Source Correlations} \label{sec:robust}
The assumption of uncorrelated sources is crucial to derive the SML method concerned in the present paper, for which the MESA algorithm is proposed. In this section, we show that the SML method is robust to source correlations, which implies robustness of MESA.



We consider the SLA case that consists of the ULA case when $M=N$. In order to show the robustness to source correlations, we will not use the statistical assumptions A1--A3 in Subsection \ref{sec:sml}. Instead, we make the following (deterministic) assumptions, where $\m{f}^o,\m{S}^o,\sigma^o$ denote the true values of the parameters.
\begin{itemize}
\item[A4:] $\m{f}^o,\m{S}^o$ are uniquely identifiable from their product $\m{Y}_{\Omega}^o=\m{A}_{\Omega}\sbra{\m{f}^o}\m{S}^o$;
\item[A5:] $\m{Y}_{\Omega} = \m{A}_{\Omega}\sbra{\m{f}^o}\m{S}^o + \m{E}$, where $\m{E}$ is random noise satisfying that $\frobn{\m{E}}^2 = NL\sigma^o>0$ and $\inf_{\m{f},\m{S}} \frobn{\m{Y}_{\Omega}- \m{A}_{\Omega}\sbra{\m{f}}\m{S}}^2$ is strictly positive;
\item[A6:] $\m{A}_{\Omega}\sbra{\m{f}}$ has full column rank in a neighborhood of $\m{f}^o$.
\end{itemize}

Assumption A4 seems necessary if no statistical assumptions are made on the source signals in $\m{S}^o$. Note that A4 implies $K<M$. Given A4, A5 is trivial given random noise $\m{E}$ since otherwise, $\m{E}$, translated by a constant vector $\m{A}\sbra{\m{f}^o}\m{S}^o$, must be in a $K$-dimensional subspace. A6 is a technical assumption ensuring that the estimates of source powers are consistent. The following proposition is a result of combining \cite[Theorem 1]{yang2016exact} and \cite[Lemma 1]{davies2012rank}.


\begin{prop} Assumptions A4 and A6 hold true if
\equ{K < \frac{\text{Spark}\sbra{\Omega} + \rank\sbra{\m{S}^o}-1}{2},}
where $\text{Spark}\sbra{\Omega}$ is defined as the smallest number of atoms in $\lbra{\m{a}_{\Omega}\sbra{f}}$ that are linearly dependent. \label{prop:assump}
\end{prop}

Our main result is stated in the following theorem.

\begin{thm} Under assumptions A4--A6 and letting $\sbra{\m{f}^*, \m{p}^*, \sigma^*}$ be the solution to the nominal SML optimization problem given by:
\equ{\begin{split}
\min_{\lbra{f_k},\lbra{p_k\geq 0},\lbra{\sigma\geq 0}}
&\left\{\ln\abs{\m{A}_{\Omega}\m{P}\m{A}_{\Omega}^H+ \sigma\m{I}}\phantom{\frac{1}{L}} \right.\\
&\quad \left.+ \frac{1}{L}\tr\sbra{\m{Y}_{\Omega}^H \sbra{\m{A}_{\Omega}\m{P}\m{A}_{\Omega}^H+ \sigma\m{I}}^{-1} \m{Y}_{\Omega}}\right\}, \end{split} \label{eq:sml3}}
we have that $\sigma^*>0$ and
\lentwo{\equa{ \lim_{\sigma^o\rightarrow0} \sigma^*
&=& 0,\label{eq:sigmaslim}\\ \lim_{\sigma^o\rightarrow0} \m{f}^*
&=& \m{f}^o, \label{eq:fslim} \\ \lim_{\sigma^o\rightarrow0} p^*_k
&=& \frac{1}{L}\norm{\m{S}^o_k}^2, \label{eq:pslim}
}}where $\m{S}^o_k$ denotes the $k$th row of $\m{S}^o$. \label{thm:robust}
\end{thm}

\begin{proof} See Appendix \ref{app:thmsigma0}.
\end{proof}

It is shown in Theorem \ref{thm:robust} that the SML method produces consistent estimates of the DOAs and source powers as the noise vanishes regardless of (spatial and temporal) source correlations, implying its robustness to (spatially) correlated or coherent sources, at least in the high SNR regime.

\begin{rem}
Theorem \ref{thm:robust} is related to \cite[Theorem 2]{yang2016enhancing} which is concerned with the problem
\equ{\min_{\m{t}} \ln\abs{\cT\m{t}+\epsilon\m{I}} + \tr\sbra{\m{Z}^H\mbra{\cT\m{t}}^{-1}\m{Z}}, \st \cT\m{t}\geq \m{0}, \label{eq:enhance}}
where $\m{Z}=\m{A}\sbra{\m{f}^o}\m{S}^o$ is noiseless and $\epsilon>0$ is a fixed small constant. In contrast to this, Theorem \ref{thm:robust} is on the noisy case in which the noise variance $\sigma$ is a variable to optimize. Another difference is that the source number is explicitly given in Theorem \ref{thm:robust}, while there is no a corresponding rank constraint on $\cT\m{t}$ in \eqref{eq:enhance}. All these differences arise due to the fact that the problem in \eqref{eq:enhance} was introduced in \cite{yang2016enhancing} as a surrogate function for the spectral sparsity of $\m{Z}$, rather than a consequence of the SML as in the present paper.
\end{rem}

\begin{rem} While the focus of this paper is on DOA estimation using SLAs, we note that Theorem \ref{thm:robust} is applicable to arbitrary linear arrays for which the sensors are not necessarily located on a regular grid. Moreover, it can easily be extended to general joint sparse recovery problems \cite{davies2012rank,wipf2007empirical}.
\end{rem}

\section{Numerical Results}\label{sec:simulation}
\subsection{Experimental Setup}
In this section, we present numerical results to illustrate the performance of the proposed MESA algorithm for DOA estimation using SLAs. In our implementation of MESA, $\m{R}_{\Omega}$ is initialized with $\widehat{\m{R}}_{\Omega}$ in general (a small scalar matrix is added if $\widehat{\m{R}}_{\Omega}$ tends to be singular), while it is initialized as an identity matrix if the ratio of the $K$th greatest eigenvalue and the smallest eigenvalue of $\widehat{\m{R}}_{\Omega}$ is smaller than a threshold (set to 5) for better performance in presence of highly correlated sources. As for the first ADMM loop, $\widehat{\boldsymbol{Y}}_{\Omega} = \widehat{\boldsymbol{R}}_{\Omega}^{\frac{1}{2}}$ and $\widehat{\boldsymbol{R}}_{\Omega}$ are used to initialize $\boldsymbol{Z}_{\Omega}$ and the corresponding principal submatrix of $\boldsymbol{Q}_3$, respectively. Other variables are initialized with zero. The outer MM loop is terminated if the relative change of the negative log-likelihood function at two consecutive iterations is lower than $10^{-5}$ or the number of iterations reaches 20. The ADMM iteration is terminated if the relative and absolute errors are below $10^{-5} $ and $10^{-4} $, respectively (see \cite[Section 3.3.1]{boyd2011distributed} for details), or a maximum number 1000 of iterations is reached. To better understand the performance of MESA, we also present its performance with a single MM iteration, termed as MESA-1. Note that the criterion of MESA-1 (when initializing $\m{R}_{\Omega}$ with $\widehat{\m{R}}_{\Omega}$) has been used in \cite{stoica2011spice,yang2014discretization} where the source number or the rank constraint is relaxed.

The methods that we use for comparison include SS-MUSIC \cite{pal2011coprime,liu2015remarks}, WLS\cite{sedighi2019asymptotically}, multiple-snapshot Newtonized orthogonal matching pursuit (MNOMP) \cite{mamandipoor2016newtonized,zhu2019multi}, reweighted atomic-norm minimization (RAM) \cite{yang2016enhancing} and maximum-likelihood estimation of low-rank Toeplitz (MELT) \cite{babu2016melt}. SS-MUSIC is a popular CBA method and is implemented with forward-backward SS and root-MUSIC. WLS is the only asymptotically efficient algorithm prior to this work when more uncorrelated sources than sensors are present. It is initialized with SS-ESPRIT following from \cite{sedighi2019asymptotically}. MNOMP is a greedy algorithm for the deterministic ML and does not require a complex initialization. RAM tries to minimize the number of sources subject to data fidelity. MNOMP and RAM do not make statistical assumptions on the sources and cannot localize more sources than sensors. RAM requires the noise power rather than the source number. MELT solves the same SML problem as MESA but is usable only for ULAs. We also compare with the CRB that is computed following from \cite{liu2017cramer} for uncorrelated sources, or based on a standard routine in presence of correlated sources.

All sources and noise are generated by using complex Gaussian distributions. All sources have unit powers. The signal-to-noise ratio (SNR) is defined as the ratio of the source power to noise power. The root mean squared error (RMSE) of the frequency estimates is computed as $\sqrt{\frac{1}{K}\left\|\widehat{\boldsymbol{f}}-\boldsymbol{f}^o\right\|_{2}^{2}}$ and then averaged over 200 Monte Carlo runs, where $\widehat{\boldsymbol{f}}$ is the vector of estimated frequencies. We consider three different types of SLAs consisting of a $10$-element ULA, a $6$-element MRA and an $8$-element nested array given respectively by
\begin{eqnarray}
\Omega_{\text{ULA}} &=& \lbra{1, 2, \dots, 10},\label{ula} \\
	\Omega_{\text{MRA}} &=& \lbra{1, 2, 7, 10, 12, 14},\label{mra} \\
	\Omega_{\text{Nested}} &=& \lbra{1, 2, 3, 4, 5, 10, 15, 20}.\label{nested}
\end{eqnarray}

\subsection{Convergence and Optimality}
In this subsection, we test the numerical performance of MESA in convergence and optimality by computing the negative log-likelihood function value at each outer MM iteration. To compare with MELT \cite{babu2016melt}, we consider the ULA in \eqref{ula}. As in \cite{babu2016melt}, the frequency domain $[-\frac{1}{2}, \frac{1}{2})$ is approximated by a set of $2N-1$ uniform gridding points and the true frequencies are selected from the gridding points to achieve the best performance for MELT, which though is not required in MESA. We also compare with the function value computed with the true values of parameters and denoted by ``ground truth''. While the globally optimal function value is hard to obtain, it is expected that good accuracy is achieved if the obtained function value is smaller than the ``ground truth''. The true noise power $\sigma^o$ is also fed into MELT as in \cite{babu2016melt}.

\begin{figure}[htbp]
	\centering
	\includegraphics[width=8cm]{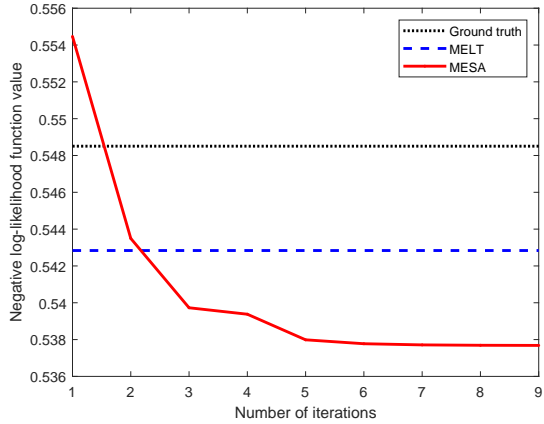}\\
	\caption{Negative log-likelihood function value versus the number of MM iterations of MESA. }\label{likeli}
\end{figure}

In {\em Experiment 1}, $K=3$ sources are generated with DOAs such that the frequencies are taken as the 5th, 7th and 18th gridding points of MELT. We set $\text{SNR}=10\text{dB}$ and the number of snapshots $L=100$. The curve of the function value with respect to the index of the MM iteration of MESA is plotted in Fig.~\ref{likeli}. It is seen that the function value decreases monotonically and converges in 9 iterations. MESA produces a function value smaller than the ground truth and MELT, indicated by the two horizontal dashed lines. MESA takes 534 inner iterations in total.


We tried a total number of 500 Monte Carlo runs and MESA always converges and produces a function value smaller than the ground truth. It performs better than MELT in 498 out of 500 runs.

%

\subsection{Statistical Efficiency for Uncorrelated Sources}
In this subsection, we use the MRA in \eqref{mra} and test the statistical efficiency of MESA for uncorrelated sources. RAM and MNOMP are considered only in the case of $K<M$.

\begin{figure}[htbp]
	\centering
	\includegraphics[width=8cm]{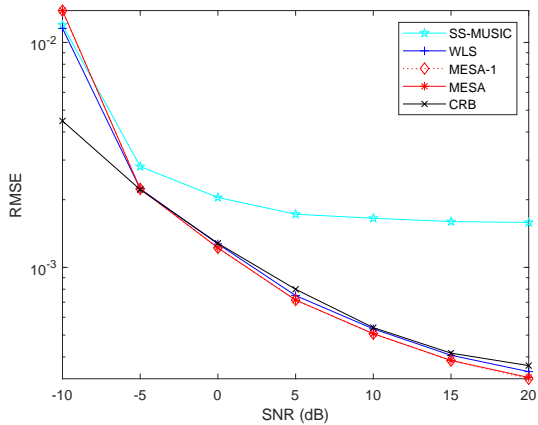}
	\caption{Frequency estimation error for $K=7$ uncorrelated sources using the MRA in \eqref{mra}, with $L=200$.}   	\label{rmse_snr}
\end{figure}

In {\em Experiment 2}, We consider $K=7$ sources with DOAs satisfying that the frequencies are taken in $\lbra{-0.43, -0.28, -0.21, -0.05, 0.1, 0.26, 0.42}$. We fix the number of snapshots $L=200$ and vary the SNR from $-10$ to $20$ dB. Our simulation results are presented in Fig.~\ref{rmse_snr}. It is seen that WLS and and MESA attain the CRB as $\text{SNR}\geq -5\text{dB}$, while SS-MUSIC always produces an error greater than the CRB. It is interesting to note that the results of MESA-1 and MESA are almost indistinguishable. In this case, the sample covariance is a good estimate of the data covariance and a single outer loop of MESA suffices to produce an accurate estimate.

%

\begin{figure}[htbp]
\centering
     \includegraphics[width=8cm]{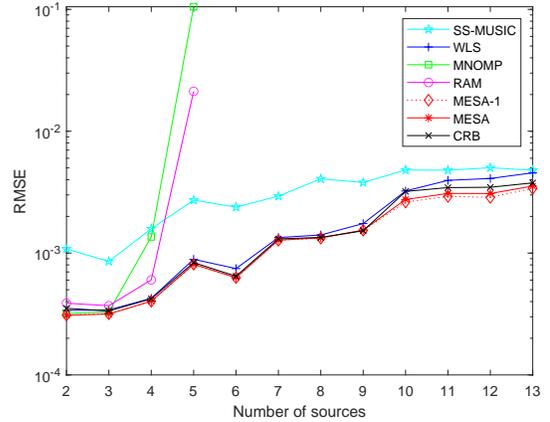}\\
     \caption{Frequency estimation error for $K\in\lbra{2,\dots,13}$ sources using the MRA in \eqref{mra}, with $L=100$ and SNR$=10$dB.}\label{rmse_k}
	\end{figure}

In {\em Experiment 3}, we fix $\text{SNR}=10\text{dB}$, $L=100$ and vary $K$ from 2 to $N-1=13$. The $K$ sources are generated with $f_k = -0.44 + 0.98(k-1)/K$, $k=1,\dots,K$. Our simulation results are presented in Fig.~\ref{rmse_k}. Again, WLS and MESA (and MESA-1) attain the CRB or even better whenever the number of sources is smaller or greater than the number of sensors, while SS-MUSIC cannot. NMOMP and RAM can accurately localize only a small number of sources, as expected.

\begin{figure}[htbp]
\centering
     	\includegraphics[width=8cm]{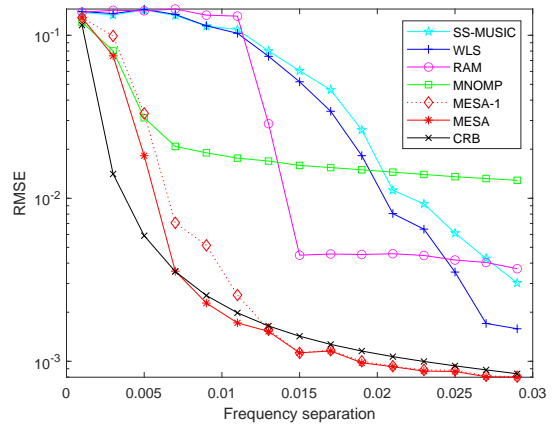}
     	\caption{Frequency estimation error for $K=3$ uncorrelated sources, two of which are separated by $\delta$, with $L=100$ and SNR $=10$dB.}      	\label{resolution}
\end{figure}

In {\em Experiment 4}, we consider $K=3$ sources with frequencies given by $\lbra{-0.2, 0.1, 0.1+\delta}$ and vary $\delta\in\lbra{0.001, 0.003,\dots, 0.029}$. We fix the number of snapshots $L=100$ and $\text{SNR}=10\text{dB}$. It is seen in Fig.~\ref{resolution} that MESA attains the CRB for very closely located sources and thus has a higher resolution than the other methods. MNOMP fails to resolve the closely located sources. A gap is shown between MESA and MESA-1, implying that the MM iterations of MESA are useful to improve the resolution.




\subsection{Robustness to Source Correlations}
\begin{figure}[htbp]
	\centering
	\includegraphics[width=8cm]{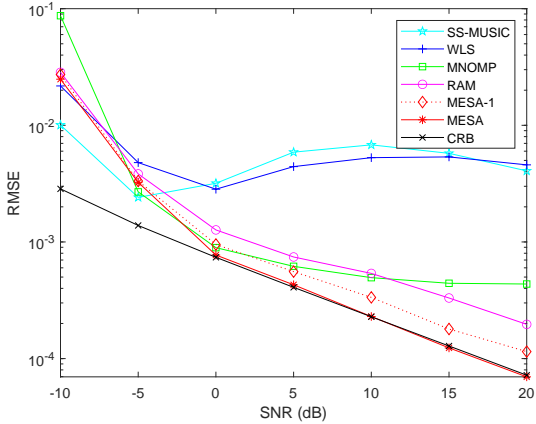}\\
	\caption{Frequency estimation error for $K = 3$ sources using the nested array in \eqref{nested}, where the first two sources are coherent.}\label{coh_nested_rmse_snr}
\end{figure}

In {\em Experiment 5}, we consider $K=3$ sources with frequencies in $\lbra{-0.2, -0.1, 0.2}$, where the first two sources are coherent with the correlation coefficient $e^{i\pi/3}$ (that can be changed to any other value on the unit circle). We use the nested array in \eqref{nested} for DOA estimation to make sure that the assumptions of Theorem \ref{thm:robust} are satisfied, which can be verified according to Proposition \ref{prop:assump} since the source matrix $\m{S}$ has rank 2 and $\text{Spark}\sbra{\Omega}\geq 6$ by noting that the nested array contains a 5-element ULA. We fix $L=100$ and vary the SNR from $-10$ to $20$dB. Our numerical results are presented in Fig.~\ref{coh_nested_rmse_snr}. It is seen that MESA has stable performance when the SNR is above 0dB, which is consistent with Theorem \ref{thm:robust}. Remarkably, MESA attains the CRB as in the case of uncorrelated sources. In contrast to this, SS-MUSIC and WLS do not have the same robustness as MESA. Satisfactory performance is also obtained by MNOMP and RAM. In this case, MESA performs better than MESA-1 since the solution to $\m{R}_{\Omega}$ is significantly different from the sample covariance and an accurate initialization is unavailable.

\begin{figure}[htbp]
	\centering
	\includegraphics[width=8cm]{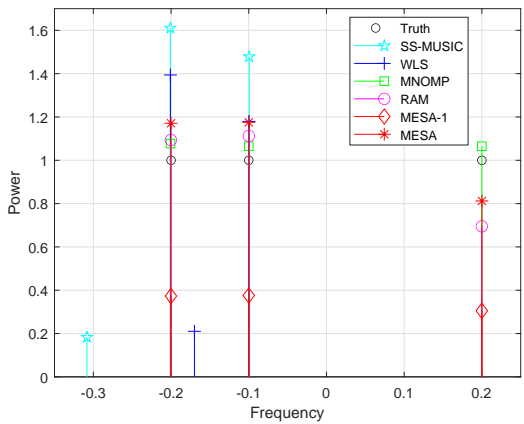}\\
	\caption{Results of one Monte Carlo run in Fig.~\ref{coh_nested_rmse_snr} at SNR $=20$dB, where the first two sources are coherent.} \label{coh_example}
\end{figure}

We present in Fig.~\ref{coh_example} results of one Monte Carlo run of the previous experiment (at $\text{SNR}=20\text{dB}$). It is seen that MESA can accurately estimate the frequencies/DOAs and the source power, validating Theorem \ref{thm:robust}. Interestingly, SS-MUSIC and WLS can accurately localize the two coherent sources but mislocate the third source that is uncorrelated with the other two. The reason underlies this behavior needs further investigation.

\begin{figure}[htbp]
	\centering
	\includegraphics[width=8cm]{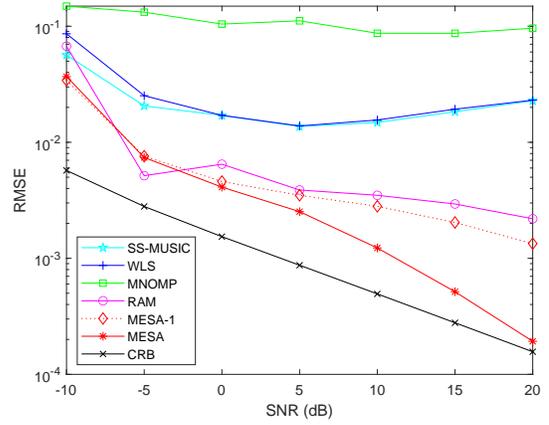}\\
	\caption{Frequency estimation error for $K = 3$ sources using the MRA in \eqref{mra}, where the first two sources are coherent.}\label{coh_mra_rmse_snr_k=3}
\end{figure}

In {\em Experiment 6}, we repeat {\em Experiment 5} by changing the nested array to the MRA in \eqref{mra}. In this case, it is difficult to verify the assumptions of Theorem \ref{thm:robust}. We present our results in Fig.~\ref{coh_mra_rmse_snr_k=3}. It is seen that all algorithms are affected to a larger extent by the source correlation as compared to the nested array case presented in Fig.~\ref{coh_nested_rmse_snr}. Differently from SS-MUSIC and WLS, MESA remains to be robust to source correlations. MNOMP has a poor performance in this case with a small array size.

\begin{figure}[htbp]
	\centering
	\includegraphics[width=8cm]{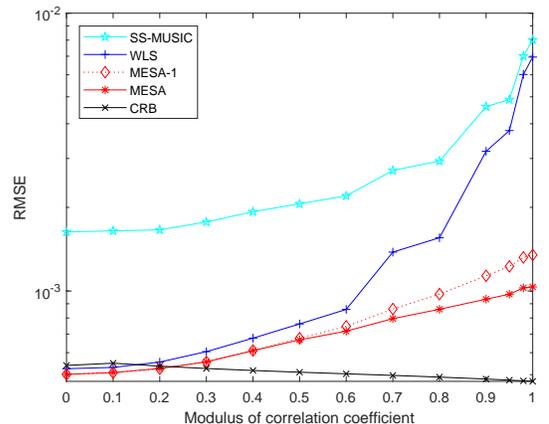}\\
	\caption{Frequency estimation error for $K = 7$ sources using the 6-element MRA in \eqref{mra}, where the first and the fourth sources are correlated.}\label{coh_rmse_CorrCoef}
\end{figure}

In {\em Experiment 7}, we repeat {\em Experiment 2} by fixing $\text{SNR}=10\text{dB}$ and letting the first and the fourth sources to be correlated with a correlation coefficient $\rho e^{i\pi/4}$, where the modulus $\rho$ changes from 0 (uncorrelated) to 1 (coherent). MNOMP and RAM are not usable since more sources than sensors are present. It is seen in Fig.~\ref{coh_rmse_CorrCoef} that the performance of all methods becomes worse as the correlation increases. In contrast to a steady performance loss of MESA, a sharp loss is shown for SS-MUSIC and WLS in the regime of highly correlated sources. As $\rho=1$, in fact, SS-MUSIC and WLS mislocate at least one of the sources in over $10\%$ of the Monte Carlo runs, while MESA always accurately localize the sources.

%

To sum up, we have shown by numerical results that MESA can statistically efficiently localize more uncorrelated sources than sensors and has robust performance in the presence of correlated and coherent sources. This makes it unique among existing coarray-based methods tailored for uncorrelated sources and sparse methods usable for a small number of sources. MESA-1 is a good accelerated approximation of MESA in general, while the latter has improved resolution and robustness. More simulation results can be found in our conference papers \cite{yang2021maximum,chen2022localizing}.

%

\section{Conclusion}\label{Conclusion}
In this paper, we showed that more sources than sensors can be localized using proper SLAs without sacrificing robustness to source correlations. This is realized by studying the robustness property of the ML method derived under the assumption of uncorrelated sources and proposing the MESA algorithm for the ML method based on elegant problem reformulations. Extensive numerical results are provided that validate our theoretical findings and demonstrate superior performance of MESA in terms of statistical efficiency, resolution and robustness to highly correlated sources as compared to state-of-the-art algorithms.

It is shown in this paper that MESA can localize more sources than sensors even in presence of highly correlated or coherent sources. A theoretical understanding of this behavior is a future work. Moreover, both algorithm-dependent and -independent analyses are of interest to investigate how the number of localizable sources leverages with source correlations. For a particular algorithm, it is shown that using the assumption of uncorrelated sources does not necessarily contradict with its robustness to correlated sources. Therefore, it is of great interest to investigate their robustness for both existing and future algorithms proposed with the uncorrelated setup. It is also shown that the array geometry is another factor affecting the robustness to source correlations. It is interesting to take the robustness into consideration for future array geometry design.

%

\appendix
\subsection{Proof of \eqref{Z update}} \label{app:updateZ}
To show \eqref{Z update}, it suffices to show that
\begin{equation}\label{z_0}
\left(1-\frac{\beta}{\left\|\m{L}\right\|_{\mathrm{F}}}\right)_{+} \m{L}=\arg \min _{\check{\boldsymbol{Z}}} \beta\|\check{\boldsymbol{Z}}\|_{\mathrm{F}} +\frac{1}{2}\left\|\check{\boldsymbol{Z}}-\m{L}\right\|_{\mathrm{F}}^{2}
\end{equation}
by identifying that $\check{\boldsymbol{Z}}=\widehat{\boldsymbol{Y}}-\boldsymbol{Z}$ and $\beta = \mu^{-1}$. To this end, observe that
\begin{equation}\label{J(z)}
\begin{aligned}
g(\check{\boldsymbol{Z}}) & \triangleq \beta\|\check{\boldsymbol{Z}}\|_{\mathrm{F}}+\frac{1}{2}\left\|\check{\boldsymbol{Z}}-\m{L}\right\|_{\mathrm{F}}^{2} \\
& \geq \beta\|\check{\boldsymbol{Z}}\|_{\mathrm{F}}+\frac{1}{2}\left(\|\check{\boldsymbol{Z}}\|_{\mathrm{F}}-\left\|\m{L}\right\|_{\mathrm{F}}\right)^{2} \\ &=\frac{1}{2}\left(\|\check{\boldsymbol{Z}}\|_{\mathrm{F}}-\left\|\m{L}\right\|_{\mathrm{F}}+\beta\right)^{2}+\beta\left\|\m{L}\right\|_{\mathrm{F}}-\frac{1}{2} \beta^{2},
\end{aligned}
\end{equation}
where the equality is achieved if $ \check{\boldsymbol{Z}} $ has the sign of $ \boldsymbol{L} $. Since the last expression is minimized if $ \| \check{\boldsymbol{Z}}\|_{\mathrm{F}}=( \| {\boldsymbol{L}}\|_{\mathrm{F}}-\beta)_+ $, the overall function $ g( \check{\boldsymbol{Z}}) $ is therefore minimized at
\begin{equation}\label{z check}
\check{\boldsymbol{Z}}=\left(\left\|\m{L}\right\|_{\mathrm{F}}-\beta\right)_{+} \operatorname{sgn}\left(\m{L}\right)=\left(1-\frac{\beta}{\left\|\m{L}\right\|_{\mathrm{F}}}\right)_{+} \m{L},
\end{equation}
completing the proof.


\subsection{Proof of Theorem \ref{thm:robust}} \label{app:thmsigma0}
We first show the following lemma.
\begin{lem} Assume $\m{A}=\mbra{\m{a}_1,\dots,\m{a}_K}$ and that $\m{P}$ is positive-definite diagonal. Then, it holds for any $\sigma>0$ and $k=1,\dots,K$ that
\equ{\m{a}_k^H\sbra{\m{A}\m{P}\m{A}^{H}+ \sigma\m{I}}^{-1}\m{a}_k< p_k^{-1}. \label{eq:pkm11} }
If, further, $\m{A}$ has full column rank, then
\equ{\lim_{\sigma\rightarrow0} \m{a}_k^H\sbra{\m{A}\m{P}\m{A}^{H}+ \sigma\m{I}}^{-1}\m{a}_k
= p_k^{-1}.\label{eq:pkm12}}
\label{lem:sigma0}
\end{lem}
\begin{proof} Note that the matrix
$\begin{bmatrix} p_k^{-1} & \m{a}_k^H \\ \m{a}_k & \m{A}\m{P}\m{A}^{H}+ \sigma\m{I} \end{bmatrix} $ is positive definite since so is $p_k^{-1}$ and its Schur complement
\equ{\m{A}\m{P}\m{A}^{H}+ \sigma\m{I} - p_k\m{a}_k \m{a}_k^H \geq \sigma\m{I}. }
Consequently, the Schur complement regarding $\m{A}\m{P}\m{A}^{H}+ \sigma\m{I}$ is positive, yielding \eqref{eq:pkm11}.

Without loss of generality, we next show \eqref{eq:pkm12} for $k=1$. For any $\epsilon>0$, $\diag\sbra{-\epsilon, p_2,\dots,p_K}$ is indefinite and so is $\m{A}\diag\sbra{-\epsilon, p_2,\dots,p_K}\m{A}^H$ since by assumption $\m{A}$ has full column rank. It follows that for any
\equ{0<\sigma < -\lambda_{\text{min}} \sbra{\m{A}\diag\sbra{-\epsilon, p_2,\dots,p_K}\m{A}^H},}
the matrix
\equ{\begin{split}
&\m{A}\m{P}\m{A}^H + \sigma\m{I} - \sbra{p_1+\epsilon}\m{a}_1\m{a}_1^H \\
&= \begin{bmatrix}\m{A}\diag\sbra{-\epsilon, p_2,\dots,p_K}\m{A}^H \end{bmatrix} + \sigma\m{I} \end{split} \label{eq:schurnew}}
is indefinite. Hence, the matrix
$\begin{bmatrix} \sbra{p_1+\epsilon}^{-1} & \m{a}_1^H \\ \m{a}_1 & \m{A}\m{P}\m{A}^{H}+ \sigma\m{I} \end{bmatrix}$
is indefinite by observing that the matrix in \eqref{eq:schurnew} is the Schur complement regarding $\sbra{p_1+\epsilon}^{-1}>0$. Since $\m{A}\m{P}\m{A}^{H}+ \sigma\m{I}$ is positive definite, consequently, its Schur complement must be negative, i.e.,
\equ{\sbra{p_1+\epsilon}^{-1} - \m{a}_1^H\sbra{\m{A}\m{P}\m{A}^{H}+ \sigma\m{I}}^{-1}\m{a}_1
< 0,}
which combined with \eqref{eq:pkm11} yields that
\equ{\sbra{p_1+\epsilon}^{-1} < \m{a}_1^H\sbra{\m{A}\m{P}\m{A}^{H}+ \sigma\m{I}}^{-1}\m{a}_1
< p_1^{-1},}
of which a direct consequence is \eqref{eq:pkm12}.
\end{proof}

We are ready to prove Theorem \ref{thm:robust}. For notational simplicity, we omit the subscript $\Omega$ in $\m{Y}_{\Omega}, \m{A}_{\Omega}$ hereafter and write them as $\m{Y}, \m{A}$ without ambiguity. It follows from Lemma 5 and Lemma 4 in \cite{yang2015gridless} that
\equ{\begin{split}
&\tr\sbra{\m{Y}^H \sbra{\m{A}\m{P}\m{A}^H+ \sigma\m{I}}^{-1} \m{Y}} \\
&= \min_{\m{Z}} \tr\sbra{\m{Z}^H\mbra{\m{A}\m{P}\m{A}^H}^{-1}\m{Z} } + \sigma^{-1}\frobn{\m{Y}-\m{Z}}^2\\
&= \min_{\m{S}} \tr\sbra{\m{S}^H\m{P}^{-1}\m{S} } + \sigma^{-1}\frobn{\m{Y}-\m{A}\m{S}}^2, \end{split}}
which can also be shown directly. Define
\equ{\begin{split}\cL\sbra{\m{f}, \m{p}, \sigma,\m{S}}
&= \ln\abs{\m{A}\m{P}\m{A}^H+ \sigma\m{I}} + \frac{1}{L}\tr\sbra{\m{S}^H\m{P}^{-1}\m{S}} \\
&\quad + \frac{1}{L\sigma}\frobn{\m{Y} - \m{A}\m{S}}^2. \end{split} \label{eq:cLnew} }
It follows that
\equ{\sbra{\m{f}^*, \m{p}^*, \sigma^*,\m{S}^*} = \argmin_{\m{f}, \m{p}, \sigma,\m{S}} \cL\sbra{\m{f}, \m{p}, \sigma,\m{S}},}
and we let $\cL^*$ denote the optimal value.

We first show $\sigma^*>0$. It suffices to show that
\equ{\lim_{\sigma\rightarrow 0} \cL\sbra{\m{f}, \m{p}, \sigma,\m{S}} = +\infty \label{eq:Latsigma0}}
for any $\sbra{\m{f}, \m{p}, \m{S}}$. To do so, note by \eqref{eq:cLnew} that
\equ{\begin{split}\cL\sbra{\m{f}, \m{p}, \sigma,\m{S}}
&\geq \ln\abs{\sigma\m{I}} + \frac{1}{L\sigma}\frobn{\m{Y} - \m{A}\m{S}}^2\\
&= M\ln\sigma + \frac{1}{L\sigma}\frobn{\m{Y} - \m{A}\m{S}}^2,\end{split} \label{eq:Lnewbd1}}
and the only stationary point of the lower bound above regarding $\sigma$, which is the global minimizer, is given by $\sigma = \frac{\frobn{\m{Y} - \m{A}\m{S}}^2}{NL}$ that is bounded from below by a positive number for any $\sbra{\m{f}, \m{S}}$ by Assumption A5. Consequently, the lower bound above always approaches infinity as $\sigma\rightarrow 0$, resulting in \eqref{eq:Latsigma0}.

Inserting the ground truth $\sbra{\m{f}^o, \m{p}^o, \sigma^o, \m{S}^o}$ into $\cL$ and conditioning on $\sigma^o\leq 1$, we have that
\equ{\begin{split}\cL^*
&\leq \cL\sbra{\m{f}^o, \m{S}^o, \m{p}^o, \sigma^o}\\
&= \ln\abs{\m{A}^o\m{P}^o\m{A}^{oH}+ \sigma^o\m{I}} + \frac{1}{L}\tr\sbra{\m{S}^{oH}\m{P}^{o-1}\m{S}^o} \\
&\quad + \frac{1}{L\sigma^o}\frobn{\m{Y} - \m{A}^o\m{S}^o}^2\\
&\leq (M-K)\ln\sigma^o + \sum_{k=1}^K\ln\sbra{\lambda_k\sbra{\m{A}^o\m{P}^o\m{A}^{oH}} + 1}\\
&\quad + K + M, \end{split} \label{eq:Lsupp}}
where $\lambda_k$ denotes the $k$th greatest eigenvalue.
Combining \eqref{eq:Lsupp} and the inequality
\equ{\begin{split}\cL^*
&\geq  \ln\abs{\m{A}^*\m{P}^*\m{A}^{*H}+ \sigma^*\m{I}} \geq M\ln\sigma^* \end{split} \label{eq:Lslow}}
yields that
\equ{\sigma^*\leq C{\sigma^o}^{\frac{M-K}{M}}, \label{eq:sigmasbd}}
where $C = e^{1+K/M}\prod_{k=1}^K \sbra{\lambda_k\sbra{\m{A}^o\m{P}^o\m{A}^{oH}} + 1}^{1/M}$ is a constant. A direct consequence of \eqref{eq:sigmasbd} is \eqref{eq:sigmaslim}.

Inserting the stationary point $\sigma = \frac{\frobn{\m{Y} - \m{A}\m{S}}^2}{NL}$ into the lower bound in \eqref{eq:Lnewbd1} yields that
\equ{\begin{split}\cL\sbra{\m{f}, \m{p}, \sigma,\m{S}}
&\geq M\ln\sigma + \frac{1}{L\sigma}\frobn{\m{Y} - \m{A}\m{S}}^2\\
&\geq M\ln \frac{\frobn{\m{Y} - \m{A}\m{S}}^2}{NL} + M.\end{split}}
Consequently,
\equ{\begin{split}\cL^*
&= \cL\sbra{\m{f}^*, \m{p}^*, \sigma^*,\m{S}^*}\geq M\ln \frac{\frobn{\m{Y} - \m{A}^*\m{S}^*}^2}{NL} + M.\end{split} \label{eq:Lslow2}}
Combining \eqref{eq:Lslow2} and \eqref{eq:Lsupp}, we obtain that
\equ{\frac{\frobn{\m{Y} - \m{A}^*\m{S}^*}^2}{NL} \leq Ce^{-1}{\sigma^o}^{\frac{M-K}{M}}.}
Therefore,
\equ{\lim_{\sigma^o\rightarrow 0} \m{A}^*\m{S}^* = \m{A}^o\m{S}^o, \label{eq:ASs}}
implying consistence of $\sbra{\m{f}^*,\m{S}^*}$ (up to perturbations of entries) by Assumption A4.

Finally, note by \eqref{eq:cLnew} that
\equ{\m{p}^* = \argmin_{\m{p}} \ln\abs{\m{A}^*\m{P}\m{A}^{*H}+ \sigma^*\m{I}} + \frac{1}{L}\tr\sbra{\m{S}^{*H}\m{P}^{-1}\m{S}^*},}
where all rows of $\m{S}^*$ are nonzero and $\m{A}^*$ has full column rank if $\sigma^o$ is small enough due to their consistency as $\sigma^o\rightarrow 0$ and Assumption A6.
It follows that the derivative of the above objective function with respect to $p_k$ vanishes at $p_k=p_k^*$, yielding that
\equ{\m{a}^H\sbra{f_k^*}\sbra{\m{A}^*\m{P}^*\m{A}^{*H}+ \sigma^*\m{I}}^{-1}\m{a}\sbra{f_k^*} - \frac{\norm{\m{S}_k^*}^2}{Lp^{*2}_k} = 0. \label{eq:pksequ}}
Applying Lemma \ref{lem:sigma0}, we then obtain
$\frac{\norm{\m{S}_k^*}^2}{Lp^{*2}_k} < \frac{1}{p_k^*}$
and thus,
\equ{p^{*}_k > \frac{\norm{\m{S}_k^*}^2}{L}}
is bounded from below by a universal positive number if $\sigma^o$ is small enough. We further apply the second part of Lemma 1 to obtain by \eqref{eq:pksequ} and \eqref{eq:sigmaslim} that
\equ{\begin{split} 0=
&\lim_{\sigma^o\rightarrow 0} \m{a}^H\sbra{f_k^*}\sbra{\m{A}^*\m{P}^*\m{A}^{*H}+ \sigma^*\m{I}}^{-1}\m{a}\sbra{f_k^*} - \frac{\norm{\m{S}_k^*}^2}{Lp^{*2}_k}\\
=& \lim_{\sigma^o\rightarrow 0} \frac{1}{p_k^*} - \frac{\norm{\m{S}_k^o}^2}{Lp^{*2}_k}, \end{split}}
which results in \eqref{eq:pslim} and completes the proof.


\section*{Acknowledgment}
The authors would like to thank Prof.~Jiang Zhu of Zhejiang University for providing the code of MNOMP.



\end{document}